\documentclass{svjour3}                     % onecolumn (standard format)
\smartqed  % flush right qed marks, e.g. at end of proof
\usepackage{graphicx}
\usepackage[misc]{ifsym}
\usepackage[justification=centering]{caption}

\usepackage{mathptmx}

\usepackage{latexsym}
\usepackage{amsmath}
\usepackage{color}
\begin{document}

\title{On-chip Multiphoton Entangled States by Path Identity }
%\thanks{The project is funded by National Natural Science Foundation of China, under Grant Nos.61003258, 61472165, and Science and Technology Planning Project of Guangdong Province, China, under Grant No.2013B010401018.}}
%\thanks{Grants or other notes
%about the article that should go on the front page should be
%placed here. General acknowledgments should be placed at the end of the article.}

%\subtitle{Do you have a subtitle?\\ If so, write it here}

\titlerunning{On-chip Multiphoton Entangled States by Path Identity}        % if too long for running head

\author{Tianfeng Feng \and Xiaoqian Zhang \and Yuling Tian \and Qin Feng}%etc.

\authorrunning{T.F. Feng et al.} % if too long for running head

\institute{ Tianfeng Feng(\Letter)  \and Yuling Tian \and Qin Feng \at  School of Physics and State Key Laboratory of Optoelectronic Materials and Technologies, Sun Yat-sen University, Guangzhou 510000, China\\
 \email{fengtf3@mail2.sysu.edu.cn}
 \and Xiaoqian Zhang(\Letter) \at Department of Computer Science, Jinan University, Guangzhou, 510632, China \\
 \email{zzhangxiaoqian@foxmail.com}
              %Tel.: +123-45-678910\\
              %Fax: +123-45-678910\\
                       %  \\
%             \emph{Present address:} of F. Author  %  if needed
}

\date{Received: date / Accepted: date}
% The correct dates will be entered by the editor

\maketitle

\begin{abstract}
  Multiphoton entanglement, as a quantum resource, plays an essential role in linear optical quantum information processing. Krenn et al. (PRL 118, 080401 (2017)) proposed an innovative scheme that generating entanglement by path identity, in which two-photon interference (called Hong-Ou-Mandel effect) is not necessary in experiment. However, the experiments in this scheme have strict requirements in stability and scalability, which is difficult to be realized in bulk optics. To solve this problem, in this paper we first propose an on-chip scheme to generate multi-photon polarization entangled states, including Greenberger-Horne-Zeilinger (GHZ) states and W states. Moreover, we also present a class of generalized graphs for W state (odd-number-photon) by path identity in theory. The on-chip scheme can be implemented in existing integrated optical technology which is meaningful for multi-party entanglement distribution in quantum communication networks.

% \PACS{PACS code1 \and PACS code2 \and more}
% \subclass{MSC code1 \and MSC code2 \and more}
\end{abstract}

\section{Introduction}
\label{intro}

Quantum entanglement, a kind of quantum correlation, plays an essential role in quantum information and quantum fundamental physics. The last few decades, quantum entanglement has been realized in various quantum systems such as superconducting qubits \cite{1}, ion traps \cite{2} and photons \cite{3,4}. Photons are often used to carry quantum information in quantum communication \cite{5,6,40,41,42}, quantum metrology \cite{7}, quantum simulation \cite{8,9,10} and quantum computing \cite{11,12,13,14} because of some excellent properties such as long coherence time, fast flying speed and multiple degrees of freedom (including polarization, spatial mode, frequency, and so on). It is an important task in optical quantum information processing to generate and manipulate multiphoton states. So far, the entanglement of eight-photon \cite{16}, ten-photon \cite{17} and twelve-photon \cite{18} has been experimentally realized by spontaneous parametric down-conversion of BBO crystals \cite{15} in bulk optics. In 2018, Adcock et al. realized four-photon path-encoded graph state based on photonic chips \cite{20}. These schemes \cite{16,17,18,15,20} of generating multiphoton entangled states are based on conventional probabilistic post-selection.

In 2017, Krenn et al. \cite{21} proposed a novel scheme that generating entanglement by path identity. In this scheme, two-photon interference is replaced by coherently pumping two independent nonlinear crystal, in which the path of photons in different quantum states are overlapping \cite{22}. In addition, the scheme \cite{21} is more flexible and robust than traditional methods \cite{16,17,18,15,20} to generate various multiphoton entangled states, such as Greenberger-Horne-Zeilinger (GHZ) states \cite{34}, W states \cite{37} and high dimensional quantum states \cite{14}. In fact, the existing techniques for generating two-photon entangled states belong to this kind of scheme that generating entanglement by path identity, such as two-layered I-type BBO down-conversion \cite{28} and Beam-like sandwich-structure BBO down-conversion \cite{29}. Recently, more relevant theoretical researches \cite{24,23,25,36} are developed and experiments on bulk optics \cite{26} are also carried out.

The phase stability of quantum states will decrease with the increase of the complexity of optical setup, which leads to the de-coherence of quantum states. Therefore, the experimental setup with complex optical paths is difficult to implement in bulk optics.  Compared to bulk optics, the optical paths of integrated optics are inherently stable. In this paper,we first propose an universal on-chip scheme to generate GHZ states and W states. Our on-chip scheme overcomes the drawbacks of bulk optics and is scalable. Besides, we also present a class of generalized graphs for W states (odd number) by path identity.

The rest of this paper is organized as follows. In Sect. \ref{sec:1}, we present some basic knowledge of core components of chip in detail. In Sect. \ref{sec:2}, multiphoton entangled state by path identify is shown. First, we show how to generate GHZ states by path identity on chip. Second, we propose a class of generalized graphs for odd-number W states. We also give an on-chip design of three-photon W states. Finally, we present the conclusions in Sect. \ref{sec:3}.

\vspace{-0.4cm}
\section{Preliminary}\label{sec:1}
\begin{figure}[!t]
  \centering
  \includegraphics[width=0.95\textwidth]{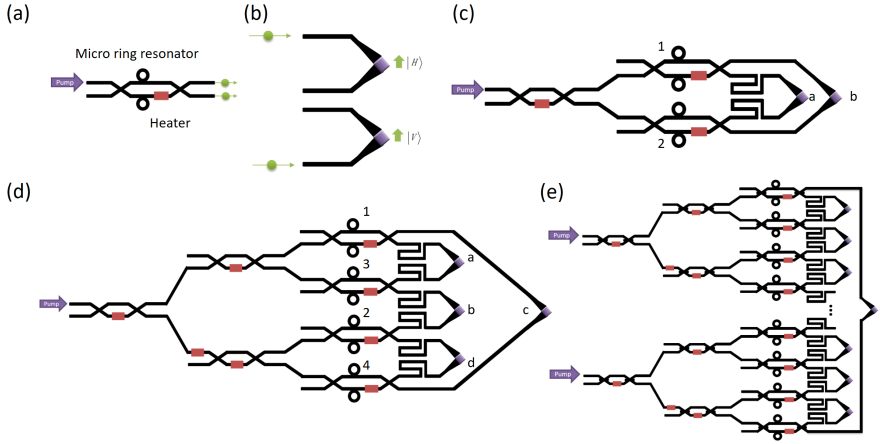}\vspace{2mm}
  \caption{(Color Online) On-chip scheme for generating multiphoton polarization entangled state by path identity. (a) Mechanism of two-photon generation. (b) Two-dimensional grating coupler. (c) Two-photon entangled state. (d) Four-photon GHZ state. (e) N-photon GHZ state.}\label{fig3}
\end{figure}

In this section, we briefly introduce the core components of chip (See Fig. \ref{fig3}(a) and (b)): two-photon source and two dimensional (2D) grating. The two-photon source has been implemented on various photonic chips \cite{19,30,31}.

In Fig. \ref{fig3}(a), a pump laser with two different wavelengths injects into silicon waveguide. Four-wave mixing effect occurs in two micro-ring resonators (or nanowires) to generate two photons with same frequency. When the micro-ring on upper path produces a pair of photons, the state is denoted as $|2,0\rangle$. Otherwise, the state is denoted as $|0,2\rangle$. Therefore, separating two photons can be realized by reversed Hong-Ou-Mandel (RHOM) effect, and this process can be expressed as

\begin{eqnarray}
 \frac{1}{\sqrt{2}}(|2,0\rangle-e^{i \Delta \varphi}|0,2\rangle)\xrightarrow{RHOM}sin(\frac{\Delta \varphi}{2})|1,1\rangle+cos(\frac{\Delta \varphi}{2})(|2,0\rangle-|0,2\rangle),
\end{eqnarray}
where relative phase $\Delta \varphi$ can be adjusted by the electric phase shifter (Heater).

Two-dimensional grating, as a fiber-chip coupling device, can convert degrees of freedom (DoF) of photon between path and polarization \cite{32,33}. The principle of 2D grating is shown in Fig. \ref{fig3}(b). When photons pass upper path, it exits with horizontal polarization, denoted as $|H\rangle$ which is equal to $|0\rangle$. Otherwise, it exits with vertical polarization, denoted as $|V\rangle$ which is equal to $|1\rangle$. After going through 2D grating, the path DoF of photons is converted to polarization DoF.

In the following, combining 2D grating, two-photon source and optical waveguide, we propose a universal on-chip scheme to generate multi-photon polarization entangled state by path identity.

\section{Multiphoton entangled state by path identify}\label{sec:2}
In this section, we give a universal scheme for GHZ state and W state respectively. Firstly, a on-chip scheme for multiphoton GHZ state is presented. Secondly, we first propose a class of generalized graphs for odd-number W state by path identity. Besides, an on-chip implementation for three-photon W state is given.

\subsection{Multi-photon GHZ  state by path identity}
As we all known, GHZ state are often used to exhibit sharp contradiction between classical physics and quantum physics \cite{34}. Recently, it is proved that GHZ state can be as units to construct cluster states in universal quantum computing \cite{38} and also can be utilized in open-destination teleportation \cite{39}. Here, we put forward an on-chip scheme to generate multiphoton GHZ state by path identity.

Fig. \ref{fig3}(c) is a chip scheme corresponding to Bell states. Source 1 produces a pair of identical photons simultaneously arriving to the upper path of two 2D gratings (the bent waveguide compensating the optical path difference of the two paths), and the photons respectively exit from port a and port b to prepare two-photon states $|HH\rangle$. Similarly for source 2, a pair of identical photons simultaneously arriving to lower path of 2D gratings, and the photons respectively exit from port a and port b to prepare two-photon states $|VV\rangle$. So far, the path information of photons has been converted to polarization information after passing 2D grating. Choosing appropriate pump power makes that more than one pair of photon can be neglected. When laser coherently pumps source 1 and source 2, two-photon state can be expressed as $|\psi\rangle=\alpha|HH\rangle+\beta|VV\rangle$, where the amplitude coefficients $\alpha$ and $\beta$ satisfy $|\alpha|^2+|\beta|^2=1$ and determined by the intensity and phase of the pump laser.

The method in Fig. \ref{fig3}(c) is applicable to multi-photon entangled state. For example, on-chip four-photon GHZ state by path identity is presented in Fig. \ref{fig3}(d). As shown in Fig. \ref{fig3}(d), a laser coherently pumps sources 1, 2, 3 and 4. The pump power is adjusted so that more than two pairs of photon can be neglected. Source 1 and source 2 simultaneously generate a pair of photons repectively. Two photons from source 1 arrive to port a and port c respectively, while two photons from source reach to port b and port d respectively. Thus, four photons are emitted in a state $|HVHH\rangle$. Similarly for source 3 and source 4, four photons are in a state $|VHVV\rangle$. Four-photon coincidences can only happen either when the two pairs come from source 1 and 2 or source 3 and 4. Therefore, four-photon state can be written as
\begin{eqnarray}
\begin{array}{l}
\displaystyle |\psi\rangle = (|H_aH_c\rangle+|V_aH_b\rangle+|V_bH_d\rangle+|V_aV_c\rangle)^2\\
\displaystyle \qquad \xrightarrow{four-fold}\frac{1}{\sqrt{2}}(|H_aV_bH_cH_d\rangle+|V_aH_bV_cV_d\rangle,
\end{array}
\end{eqnarray}
where the subscript $a, b, c, d$ indicate the path of photons respectively. In theory, the chip can be extended to N-photon GHZ state (See Fig. \ref{fig3}(e)) expressed as $|\psi\rangle=\frac{1}{\sqrt{2}}(|HHH...\rangle+|VVV...\rangle)$.

\subsection{ W state by path identity}
W state \cite{37}, as a class of Dicke states \cite{35}, is more robust to noise. The three-photon W state can be expressed as $|\psi\rangle=\frac{1}{\sqrt{3}}(|100\rangle+|010\rangle+|001\rangle)$. Even if a photon is lost in a three-photon W state, the rest of photons keep maximal entanglement. Therefore, W state has potential applications in quantum communication and quantum networks. In this section, we propose a class of generalized graphs to generate odd-number-photon W state by path identity, meanwhile we also put forward an on-chip implementation scheme for three-photon W state.

\begin{figure}[!t]
  \centering
  \includegraphics[width=1\textwidth]{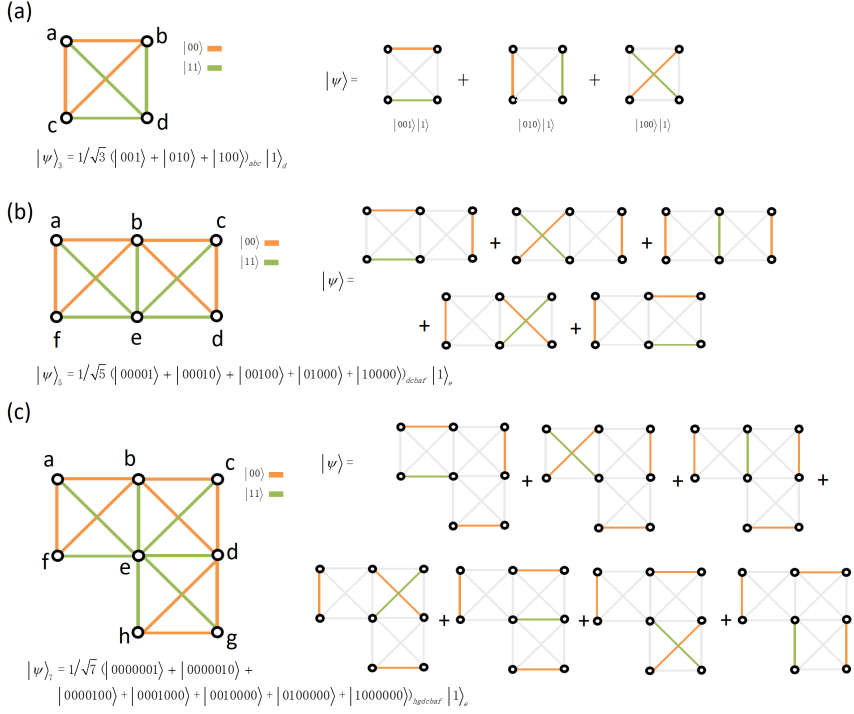}
  \caption{(Color Online) Odd-number W state by path identity. Vertices represent the path of photons and connected lines are corresponding to nonlinear crystals. Vertices connected by orange lines denotes state $|00\rangle$ and by green lines denotes state $|11\rangle$. (a) Graphs of three-photon W state. (b) Graphs of five-photon W state. (c) Graphs of seven-photon W state. }\label{fig2}
\end{figure}

For better describing complex optical experiments, Krenn et al. \cite{23} proposed a theory that maps graph theory to linear optical quantum experiments. Here, we give a graph description of odd-number W state for the first time. In linear optics, a three-photon W state can be generated by two pairs of photons (four photons) in which one photon acts as a trigger. As shown in Fig. \ref{fig2}(a), vertices a, b and c represent three photon paths of W state respectively while one photon on path d is as trigger.

The three-photon state is the superposition of three orthogonal states $|001\rangle$, $|010\rangle$ and $|100\rangle$. The three orthogonal states are respectively generated by three processes in right three graphs of Fig. \ref{fig2}(a) which corresponding to $|001\rangle|1\rangle$, $|010\rangle|1\rangle$ and $|100\rangle|1\rangle$. The first process is that a pair of photons is in state $|00\rangle$ corresponding to port a and port b while the other pair of photons ise in state $|11\rangle$ corresponding to port c and port d. When detector d fires, the three-photon state is in state $|001\rangle$. Similarly, the second process is that a pair of photons is in the state $|11\rangle$ relating to port a and port c while the other pair of photons is in state $|00\rangle$ relating to the b and d ports. When d detector fires, the three-photon state is $|010\rangle$. The third process is that a pair of photons is in state $|00\rangle$ linking to port b and port c while the other pair of photons is in state $|11\rangle$ linking to port a and port d. When detector d fires, the three-photon state is $|100\rangle$. By above three processes, the three-photon W state can be expressed as follows.
\begin{eqnarray}
\begin{array}{l}
\displaystyle |\psi\rangle_{3} = (|0_a0_b\rangle+|0_a0_c\rangle+|0_b0_c\rangle++|1_a1_d\rangle+|1_b1_d\rangle+|1_c1_d\rangle)^2\\
\displaystyle \qquad\xrightarrow{four-fold}\frac{1}{\sqrt{3}}(|0_a0_b1_c\rangle+|0_a1_b0_c\rangle+|1_a0_b0_c\rangle)|1_d\rangle.
\end{array}
\end{eqnarray}

It is similar to construct five-photon W state (six-fold) and seven-photon W state (eight-fold) via overlapping one green edge of units (See Fig. \ref{fig2}(b) and (c)). After performing these processes in Fig. \ref{fig2}(b) and (c), N-photon (N is odd number) W state can be generated, in which the W state is written as $|\psi\rangle_N=\frac{1}{\sqrt{N}}(|100\ldots 0\rangle+|010\ldots 0\rangle+|001\ldots 0\rangle+\cdots+|000\ldots 1\rangle)$. That is, we give a universal scheme for odd-number-photon W state.

\begin{figure}[!h]
  \centering
  \includegraphics[width=0.8\textwidth]{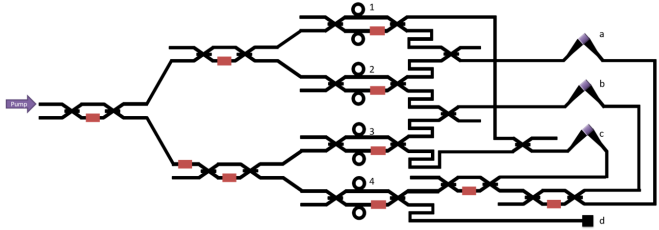}
  \caption{(Color Online) On-chip implementation of three-photon W state by path identity.}\label{fig4}
\end{figure}

Next, we analyze the generation of W state on photonic chip. In Fig. \ref{fig4}, we show a simple on-chip implementation of three-photon W state. The multimode interferometer (MMI) and Mach-Zehnder interferometer (MZI) are utilized to assist in generating three-photon W state. The source 4 produces a pair of photons, in which the one goes to port d (one-dimensional grating, we define output state of d is $|V_d\rangle$), and the other one passes through two MZI interferometers with a one-third probability goes to ports a, b, and c. That is, two-photon coincidence may occur in ports a and d, or ports b and d, or ports c and d. When source 2 and source 4 respectively generates a pair of photons at the same time. Two photons generated by source 2 reach to ports a and b respectively with a quarter probability (due to the loss of MMI). The other two photons generated by source 4 arrive to ports c and d respectively with one-third probability. After four-photon coincidence, the output states in ports a, b, c and d are $|H_aH_bH_c\rangle|V_d\rangle$. Similarly, when source 1 and source 4 respectively generates a pair of photons at the same time, the output states in ports a, b, c and d are $|H_aV_bH_c\rangle|V_d\rangle$. For sources 3 and sources 4, the output states in ports a, b, c, and d are $|V_aH_bH_c\rangle|V_d\rangle$. Therefore, we can know that the total probability of above three processes is one-twelfth. Note here, coherently pumping four two-photon sources and selecting appropriate pump power make that the cases of more than two pairs of photons can be neglected. After post-selecting four-photon coincidence, the final four-photon state can be expressed as $|\psi\rangle=\frac{1}{\sqrt{3}}(|H_aH_bV_c\rangle+|H_aV_bH_c\rangle+|V_aH_bH_c\rangle)|V_d\rangle.$ When one photon is detected on port d , three-photon W state is prepared successfully.

\section{Conclusions}\label{sec:3}
In this paper, we deeply study the entanglement by path identity and propose a universal on-chip scheme for preparing multiphoton entangled states such as GHZ states and W states. This scheme can be extended to more complicated optical network with two advantages, i.e. outstanding scalability and stability. Our on-chip scheme can be implemented in existing technology of integrated optics. Moreover, we present a class of generalized graphs for odd-number W state by path identity in theory. Our proposed multiphoton-entanglement chip can be applied to multi-party entanglement distribution in quantum communication networks.

%\vspace{-0.4cm}
\section*{Acknowledgments}
The research is funded by Project supported by the National Science Foundation of Guangdong Province, China (Grant No.2016A030312012).

%\begin{acknowledgements}
%If you'd like to thank anyone, place your comments here
%and remove the percent signs.
%\end{acknowledgements}

% BibTeX users please use one of
%\bibliographystyle{spbasic}      % basic style, author-year citations
%\bibliographystyle{spmpsci}      % mathematics and physical sciences
%\bibliographystyle{spphys}       % APS-like style for physics
%\bibliography{}   % name your BibTeX data base

% Non-BibTeX users please use
%
% and use \bibitem to create references. Consult the Instructions
% for authors for reference list style.
%\vspace{-0.4cm}

\end{document}